\documentstyle[prd,aps,preprint,epsfig,tighten]{revtex}

\def\[{\left [}
\def\]{\right ]}
\def\({\left (}
\def\){\right )}

\newcommand\be{\begin{equation}}
\newcommand\ee{\end{equation}}
\newcommand\bea{\begin{eqnarray}}
\newcommand\eea{\end{eqnarray}}

\newcommand\GeV{\,\mbox{GeV}}

\newcommand\mpl{M_{\rm Pl}}

\def\lsim{\lesssim}

\tighten

\begin{document}
\begin{titlepage}
\pagestyle{empty}
\begin{center}
\tighten
    \hfill LANCS-TH/98705 \\
    \hfill \today 
\vskip .2in
{\large \bf Allowed parameter regions for a general inflation moded 
}\footnote{
This work was supported in part by the National Nature Science Foundation of P.R.China} 
\vskip .2in
\vskip .2in
Xin He Meng

\vskip .2in

{\em CCAST,PO Box 8730, Beijing \\ }
{Department of Physics, Nankai University, Tianjin, P.R.China\\ }
{\em Department of Physics, Lancaster University, 
Lancaster, LA1 4YB, U.K.}

\end{center}

\tighten

\begin{abstract}

The early Universe  inflation is well known as a promising theory to explain the origin of large scale structure of Universe and to solve the  early universe pressing problems\cite{tur,mic}. For 
a resonable inflation model, the potential during inflation must be very flat in, at least, the direction of the inflaton. To construct 
the inflaton potential all the known related astrophysics observations should be included. For a general tree-level hybrid inflation
 potential, which is not discussed fully so far, the parameters in it are shown how to be constrained via the astrophysics data observed and to be obtained to the expected accuracy, and consistent cosmology requirements.

\end{abstract}
\end{titlepage}

\section{introduction}

For the past years, two major theories, inflation and topological defects, to the early universe problems\cite{tur} seem to have stood the test of time and one current goal is to determine which if any best fits the increasingly accumulated astrophysics data, especially the COBE's cosmic microwave background anisotropy detections, and more reasonably interprete the origin of the Universe large scale structures. They usually are regarded as mutually exclusive theories in that defects formed before a period of inflation would rapidly be diluted to such a degree during the inflationary era as to make them of little interest to cosmology. But in some inflation models inspired from particle physics considerations the formation of topological defects can be naturally obtained at the end of the inflationary period\cite{lyth1}. On the other side, as is widely supposed, the initial conditions for the successful hot big bang are set by inflation, then an adiabatic, Gaussian and more or less scale invariant density perturbation spectrum at horizon entry is predicted\cite{lyth}. Such a perturbation is generated by the vacuum fluctuation during inflation so the dazzling prospect of a window on the fundamental interactions on scales approaching the Planck energy appears. The studying of inflation paradigm will help us to understand basic physics laws in the Nature.

The inflationary Universe scenario\cite{LL2} has the universe undergoing a period of accelerated expansion, the effect being to dilute monopoles (and any other defect formed before this period) outside of the observable universe, thereby dramatically reducing their density to below the observable limits. In a homogeneous, isotropic Universe with a flat Friedmann-Robertson-Walker (FRW) metric described by a scale factor a(t), the acceleration is given via $\ddot{a}=-a(4\pi/{3\mpl^2})(\rho+3p)$ where $\rho$ is the energy density and p the pressure. Usually the energy density which drives inflation is identified with a scalar potential energy density  that is positive, and flat enough to result in an effective equation of state  $\rho\approx-p\approx V(\phi)$ satisfying the condition $\ddot{a}>0$. 

The scalar potential is associated with a scalar field known as the so-called inflaton. During the inflationary period, the inflaton potential is fairly flat in the direction the field evolves, dominating the ennergy density of the universe for a finite period of time. Over the period it evolves slowly towards a minimum of the potential either through a classical roll over or through a quantum mechanics tunnelling transition. Inflation then ends when the inflaton starts to execute decaying oscillations around its own vacuum value, and the hot Big Bang (reheating) ensues when the vacuum value has been achieved and the decay products have thermalised. Over the past  decades there have been lots of inflation models constructed and to be built. With our knowledge so far we understand that any reasonable inflation model should satisfy at least that COBE normalization, cosmology observations constraint to the spectral index and adequate inflation for consistence requirements\cite{LL2}.         
 
This paper is arranged as following. In next section we give a general comments on the properties of inflaton potential, which must satisy the COBE normalization condition . In section three we examine detailly a false vacuum dominated inflation potential model which can be regarded as a generalization of previously several fully discussed inflation models\cite{lind}. We use the slow-roll approximation to derive an analytic expression for the e-folds number N between a given epoch and the end of slow-roll inflation and derive the spectral index of the spectrum of the curvature perturbation for this model.  Confronted them with the COBE measurement of the spectrum on large scales ( the normalization), the required e-folds number N and the observational constraint  on the spectrum index over the whole range of cosmological scale, we give the  inflaton model 
an allowed region, to specify  its paremeter space by reducing its two free parameters to one .  Finally we give a discussion and conclusion on it. 

\section{general considerations on Inflation model} 
Cosmological inflation has been regarded as the most elegant solution 
to the horizon and flatness problems of the standard Big Bang 
universe\cite{LL2}.  Even though it explains  why the current 
Universe appears so homogenous and flat in a natural manner, it 
has been difficult to construct a model of inflation without a small 
parameter( fine-tunning problem).  The key point is to have a 
resonable scalar field potential either from a more underlying 
 gravity theory like effective superstring thery or from a more fundmental particle physics theory such as supergravity. In fact
to any case, one needs at least a scalar field (inflaton) that rolls 
down the potential very slowly with enough e-folds number to successfully generate a viable
inflationary scenario.
This requires the potential to be 
almost flat in the direction of the inflaton. There are lots of inflation models constructed so far.  If gravitation wave contribution is negligible, at least for the present situation, the curvature perturbation
spectrum index is the most powerful discriminator to inflation models.	In this section 
we discuss the general properties of an inflaton potential with the astrophysics considerations.
In the effective  slow-roll inflation scheme, the inflaton potential $V(\phi)$
must satisfy the flatness conditions $\epsilon\ll 1$
and $|\eta|\ll 1$, where \cite{LL2}
\bea
\epsilon &\equiv & \frac12 \mpl^2 (V'/V)^2 \\
\eta &\equiv & \mpl^2 V''/V
\eea
the prime indicates differiential with respect to $\phi$ and $\mpl = (8\pi G)^{-1/2} = 2.4\times 10^{18}\GeV$ is the reduced
Planck mass.  When these are satisfied, the time dependence of the
inflaton $\phi$ is generally given by the slow-roll expression
$3H\dot\phi = -V'$, where $H\simeq \sqrt{\frac{1}{3}\mpl^{-2} V}$ 
is the Hubble parameter during inflation.
On a given scale, the spectrum of the primordial curvature perturbation,
thought to be the origin of structure in the Universe,
is given by
\be
\delta_H^2 (k) = \frac1{150\pi^2 \mpl^4} \frac V \epsilon
\ee
The right hand side is evaluated when the relevant scale
$k$ leaves the horizon. On large scales, the COBE observation
of the cmb anisotropy corresponds to
\be
V^{1/4}/\epsilon^{1/4} = .027\mpl = 6.7 \times 10^{16}\GeV
\label{cobe}
\ee

The spectral index of the primordial curvature perturbation
is given by
\be
n-1 = 2\eta - 6\epsilon 
\ee
A perfectly scale-independent spectrum would correspond to $n=1$, 
and observation already demands $|n-1| < 0.2$. Thus $\epsilon$ 
and $\eta$ have to be $\lsim 0.1$ (barring a cancellation)
and this constraint will get tighter
if future observations, such as the near future PLANCK mission,
move $n$ closer to 1. Many models of inflation
predict that this will be the case, some giving a value of $n$ 
completely indistinguishable from 1.

Usually, $\phi$ is supposed to be charged under at least a
$Z_2$ symmetry $\phi\to-\phi$, which is unbroken during inflation.
Then $V'=0$ at the origin, and inflation typically takes place near the 
origin. As a result $\epsilon$ negligible compared with $\eta$, and
$n-1 = 2\eta\equiv 2\mpl^2 V''/V$. We assume that this is the case as in most inflation models, in 
what follows. If it is not, the
inflation  model-building, as in the slow roll approximation, is even more tricky. 

\section{A tree-level hybrid inflation  model}
In this section we will present the allowed parameter regions for a particular false vaccum-dominated potential that are not  given out before. A similar form potential with only different signs appears in a Susy model\cite{riot}.
The focused false vacuum dominated potential we consider has the usual
form of the chaotic inflationary potential \be V=V_0+m^2\phi^2/2 +
\lambda\phi^4/4\label{v}\,.\ee with all parameters positive in it, which can be regarded as a generalization of previously fully discussed inflation models\cite{lind}. Due
to symmetry considerations we discard the cubic term. Higher order
inflaton terms may appear in some Susy particle physics
models\cite{LL2,susy,dine}. There are two particular limits of false
vacuum energy inflation, according to whether the energy density is
dominated by the false vacuum energy density or by the inflaton energy
density. We assume the former in our case as preference in the slow-roll approximation \cite{lyth}.
With the COBE normalization $ V^{1/4}/\epsilon^{1/4} = .027\mpl$ and
$\epsilon \ \equiv\  \frac12 \mpl^2 (V'/V)^2 $, in our case the false
vacuum energy density \be V_0\approx 0.3^4 \mpl^2
(m^2\phi_1+\lambda\phi_1^3)^{2/3}/2^{1/3} \,.
 \ee
 where $\phi_1$ is
the inflaton  value when COBE scale leaves the horizon.
To reduce free parameters number we define $y_1=m^2/\phi_1^2 $. 
By cosmology observations to the power spectral index
constraint $|n-1|/2=\eta= \mpl^2 V''/V<0.1$ and if we take the nowadays
observation value upper limit 0.1 as a potentially changing parameter x, we have 
\be \eta=\mpl^2(m^2+3\lambda\phi^2)/V_0 < x\,.\ee
Taking equation (7) into relation (8) and using our definition for $y_1$ we candefine a function of $y_1$ as 
\be f(y_1)=\frac{y_1+3\lambda}{(y_1+\lambda)^{2/3}} < x  0.3^4/2^{1/3}
\label{7}\,.\ee
In this expression, the parameter $x
$ as an observation input runs from nowadays 0.1 to the hopfully 0.01 by the  future planned  PLANK satellite mission.
With obviously
$y_1+\lambda < y_1+3\lambda$ and relation (9) we have
\be y_1+\lambda < x^3 0.027^4/2\,.\ee 
which can give the allowed parameters regions for $y_1$ and $\lambda$ with various observed or to be obtanined parameter x 
 values.

With the certain e-folds number constraint for overcoming horizon and flatness problems 
\be N=|\int_{\phi_1} ^{\phi_c}\frac{ V}{V'}{ d\phi}|/\mpl^2\,.\ee
where $\phi_c$ is the inflaton at the end of inflation. Insert the portential form (6) and COBE normlization (7) 
 we will get another relation for a  reduced parameter  defined as 
\be y_c=m^2/\phi_c^2=(\lambda+y_1)exp(Ny_1 10^3/3.6\times(\lambda+y_1)^{-2/3})-\lambda\,.\ee
By which and relation (9) we can have the allowed parameter regions with N=50 and x=0.1 as today's observations required values in fig.1. We can see the limit cases are consistent with the previous results\cite{lind}.
We also can get directly  from (12) relation curves of the reduced parameter $y_c$ with parameter $\lambda$ as fig.2. It is clearly shown that the reduced parameter $y_c$ is approximatly linear to $\lambda$ when the expenatial value is around 1.
\begin{figure}[h!]
\begin{center}
\epsfig{file=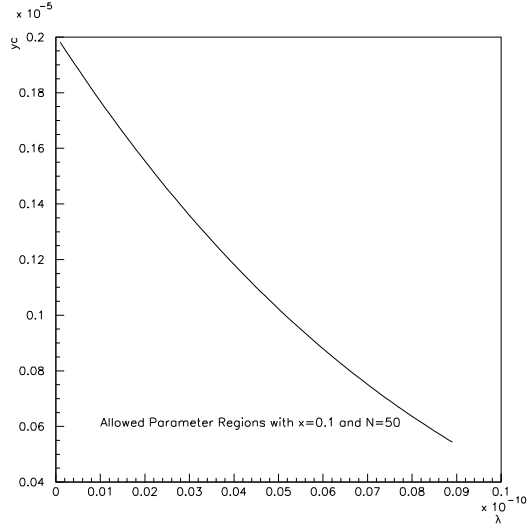,width=0.45\textwidth}
\caption{ Allowed parameter regions for $y_c$ and  $\lambda$}
\label{s}
\end{center}
\end{figure}

\begin{figure}[h!]
\begin{center}
\epsfig{file=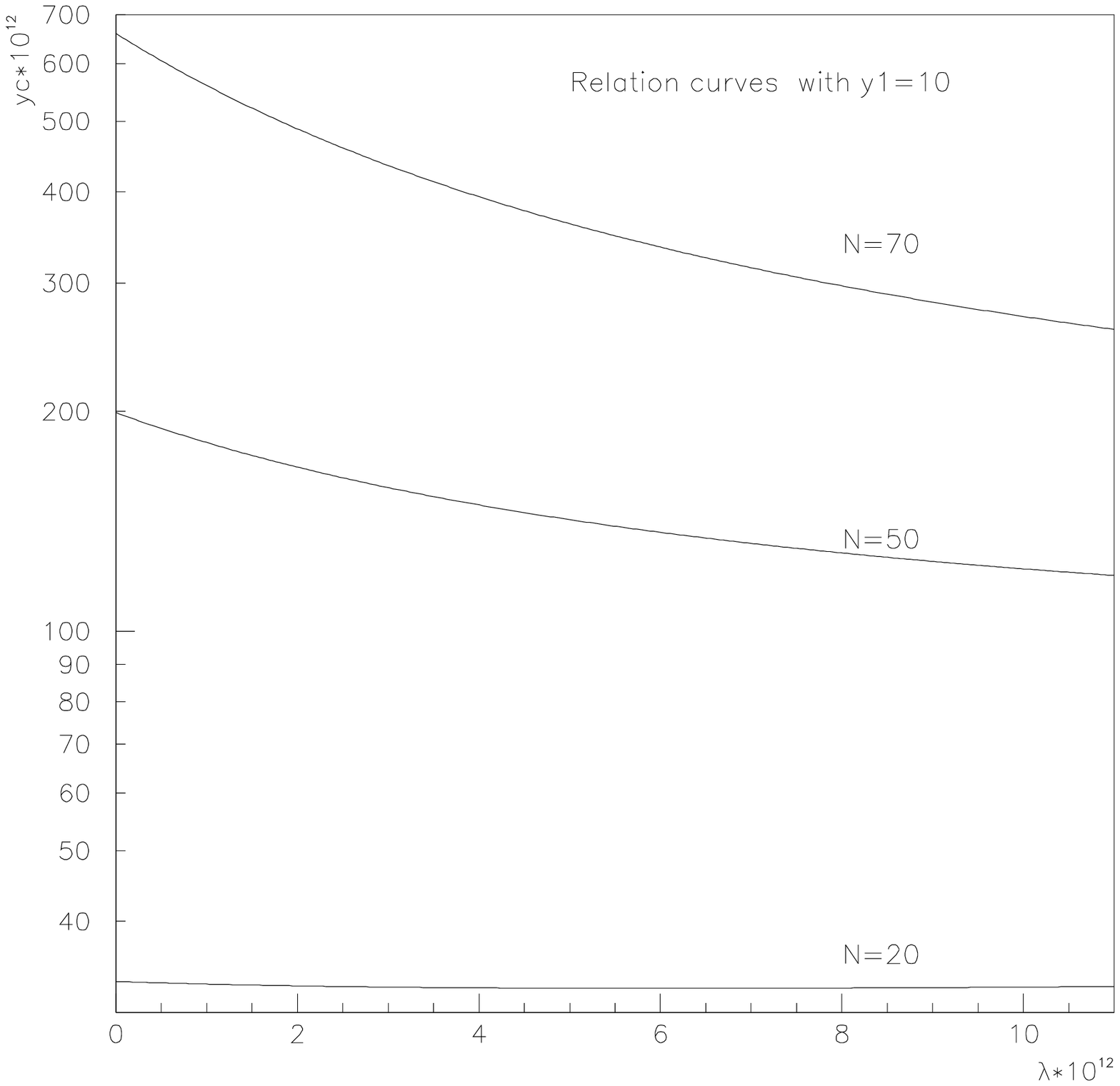,width=0.45\textwidth}
\epsfig{file=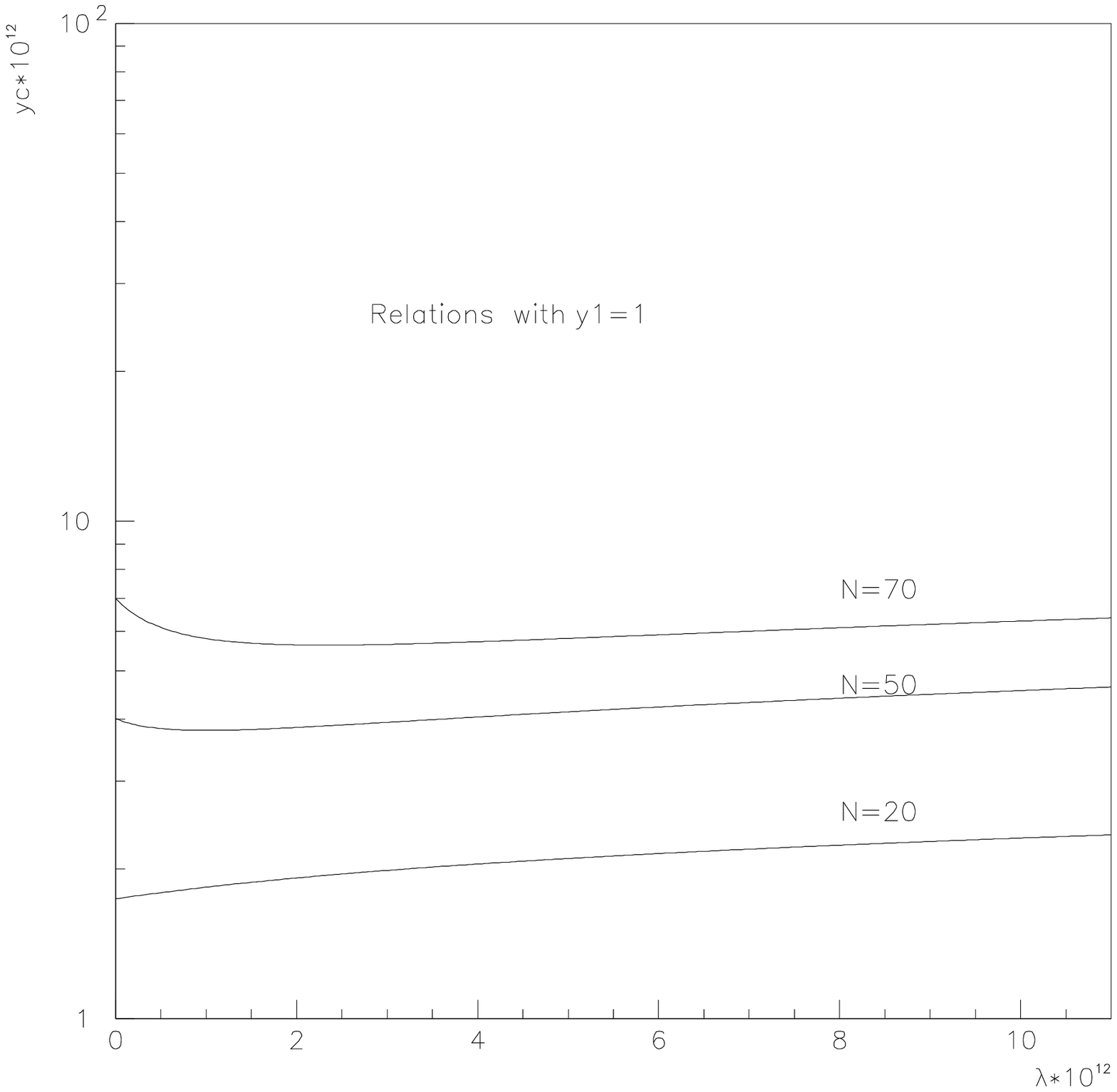,width=0.45\textwidth}
\caption{ $y_c$ as a function of $\lambda$}
\label{s}
\end{center}
\end{figure}

Taking relations (9) and (10) with equation (12) into account we can get the constraint relation,  approximatedly allowed regions to reduced parameters $y_c$ and $\lambda$
\be y_c+\lambda<(y_1+\lambda)exp(1000 N/3.608\times (y_1+\lambda)^{1/3})<2.657
\times10^{-7} x^3 exp(1.6518Nx)\ ,\ee 
By 
which we can also have the approximatedly allowed parameter regions for  $\lambda$ vs $y_c$ with x from 0.1 to future possibly 0.01 as fig.\ref{s} which is consistent with fig.1.
\begin{figure}[h!]
\begin{center}
\epsfig{file=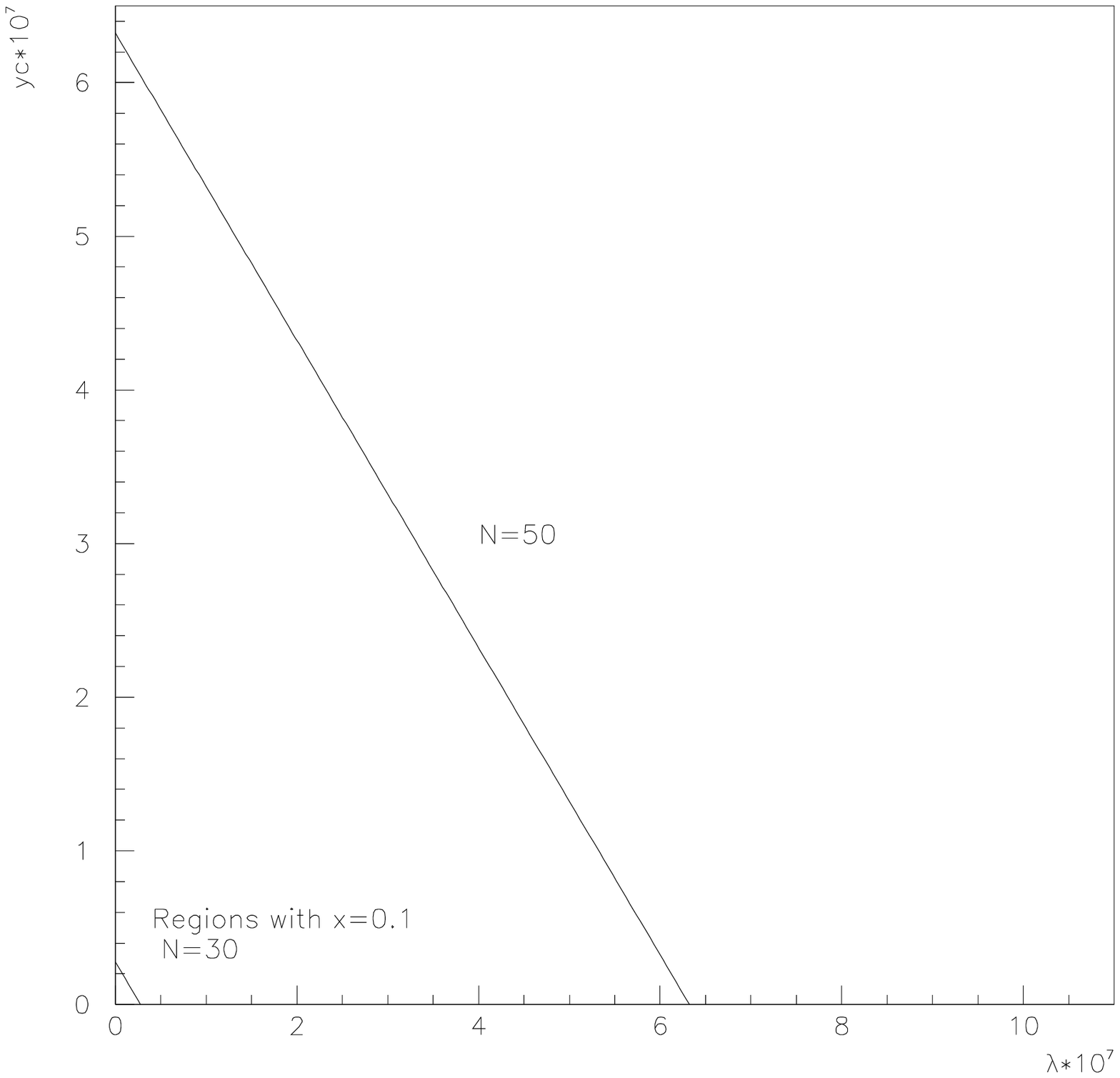,width=0.45\textwidth}
\epsfig{file=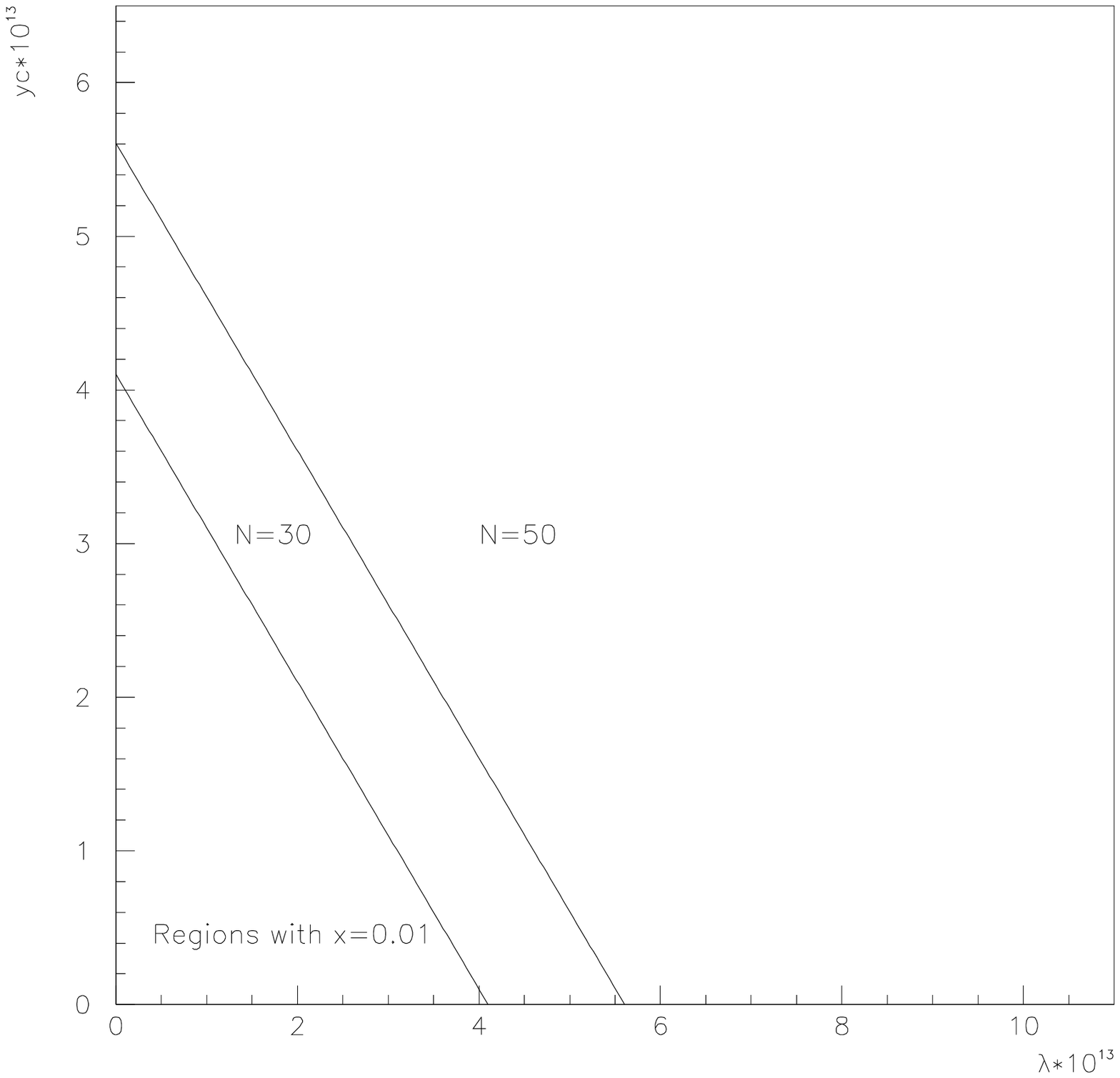,width=0.45\textwidth}
\caption{ Allowed parameter regions with different N and x values}
\label{s}
\end{center}
\end{figure}

 If we put the spectral expression and the e-folding expression together we can find 
\be |n-1|=N^{-1}2^{1/6}(1+3\lambda/y_1)ln(1+\frac{y_c-y_1}{\lambda+y_1})\,.\ee
qualitively that roughly is \be |n-1|\propto N^{-1}\,.\ee which implies these two constrints have intrinsic connection. When $|n-1|<0.1$ as today's cosmological observations present then $N\gtrsim 10$. If $|n-1|<0.01$ as the soon satellite missions by MAP and PLANCK on design  hopefully to give, then the $N\gtrsim100$. 

\section{discussion and conclusion}
Models of inflation driven by a false vacuum are mainly different from true vacuum cases in their no zero false vacuum energy density, which are simple but also can reflect the astrophysics obervations. We discuss a regular chaotic inflation model(not a toy model) here to show how to constraint its parameters when confronting data and cosmology consistence requirements, and give several new  parameter relations and the allowed regions, which can be used as a prototype model for the two planned Microwave Anisotropy Probe (MAP) and PLANCK satellite missions tests. The results we have obtained based on a seemly reasonable assumpation that the spectral index constraint we concentrate on is naturally satisfying the flatness conditions. Otherwise the slow roll approximation is  not appliable.

The origin of this tree-level hybrid inflation model or its more complicated extentions may arise from some kind of supergravity models which is generally cosidered as the appropriate framework for a description of the fundamental interactions, and in particular for the description of their scalar potential. Here we only study the essence of them in order to get more viable inflation models from supergravity theories. No matter what kind of theoretical model to be built it must satisfy at least the above observations, especially the spectral index constraint at x=0.01.The parameter space then in our case is very tiny that asks us to build the inflation models from a more natural way to avoid the fine-tunning problem, which is a chanlenge facing us.  Within next few years with the  dramatic increase in the variety and accuracy of cosmological observation data, like the measurements of temperature anisotropies in cosmic microwave background at the accuracy expected from MAP and PLANCK, it is possible for us to discriminate among inflation models.

\vskip 1cm
\underline{Acknowledgements}:
The author is very grateful to D.Lyth for suggestion and many discussions as well as comments to this work.
He also thanks D.Lyth, L.Covi and L.Roszkowski
 for many helps and hospitality extended to him during this work.
This work is partially supported by grants from China Education Committee and from NSF of China. 

\def\NPB#1#2#3{Nucl. Phys. {\bf B#1}, #3 (19#2)}
\def\PLB#1#2#3{Phys. Lett. {\bf B#1}, #3 (19#2) }
\def\PLBold#1#2#3{Phys. Lett. {\bf#1B} (19#2) #3}
\def\PRD#1#2#3{Phys. Rev. {\bf D#1}, #3 (19#2) }
\def\PRL#1#2#3{Phys. Rev. Lett. {\bf#1} (19#2) #3}
\def\PRT#1#2#3{Phys. Rep. {\bf#1} (19#2) #3}
\def\ARAA#1#2#3{Ann. Rev. Astron. Astrophys. {\bf#1} (19#2) #3}
\def\ARNP#1#2#3{Ann. Rev. Nucl. Part. Sci. {\bf#1} (19#2) #3}
\def\MPL#1#2#3{Mod. Phys. Lett. {\bf #1} (19#2) #3}
\def\ZPC#1#2#3{Zeit. f\"ur Physik {\bf C#1} (19#2) #3}
\def\APJ#1#2#3{Ap. J. {\bf #1} (19#2) #3}
\def\AP#1#2#3{{Ann. Phys. } {\bf #1} (19#2) #3}

\end{document}